\documentclass[12pt,twoside,a4paper]{article}
\usepackage{axodraw}
\usepackage{epsfig}
\usepackage{amsmath}
\author{Jan-Markus Schwindt and Christof Wetterich}
\date{}

\title{Chiral tensor fields and spontaneous breaking of Lorentz symmetry}

\begin{document} 
\maketitle 

\centerline{\small\it Institut f\"ur Theoretische Physik, 
Philosophenweg 16, 69120 Heidelberg, Germany}
\centerline{\small\it E-mail: Schwindt@thphys.uni-heidelberg.de,
C.Wetterich@thphys.uni-heidelberg.de}
\vspace{0.7cm}

\begin{abstract}
Antisymmetric tensor fields interacting with quarks and leptons have
been proposed as a possible solution to the gauge hierarchy problem.
We compute the one-loop beta function for a quartic self-interaction
of the chiral antisymmetric tensor fields. 
Fluctuations of the top quark drive the corresponding running coupling
to a negative value as the renormalization scale is lowered. This may
indicate a non-vanishing expectation value of the tensor field, and thus a
spontaneous breaking of Lorentz invariance. Settling this issue will need the 
inclusion of tensor loops. 

\end{abstract}

\section{Introduction}

A recent proposal for a solution to the gauge hierarchy problem replaces the Higgs doublet
of the electroweak standard model by chiral tensor fields \cite{cw2}. 
Symmetries forbid a local mass term for these chiral tensors.
The chiral couplings
between the tensor fields and quarks and leptons are asymptotically free, such that they
grow large at a new non-perturbative scale $\Lambda_{ch}$. Similar to
the confinement scale $\Lambda_{QCD}$ in strong interactions, this scale is
exponentially small as compared to the unification scale. It was advocated that close to the 
scale $\Lambda_{ch}$ a bound state of top-antitop forms, due to the strong attractive
interaction mediated by the tensor fields. If this collective state condenses, the
electroweak gauge symmetry is spontaneously broken and the W and Z bosons as well
as the quarks and leptons acquire a mass. The role of the fundamental Higgs doublet in the
standard model is taken by the ``composite'' top-antitop condensate \cite{tc1,tc2}.
First phenomenological studies show no contradiction with present observations
\cite{cw2,cw1,cw4}. 
In particular, the flavour and CP violation is described by the CKM matrix \cite{ckm1,ckm2}, 
as in the standard model.

The free theory for chiral tensor fields is not classically stable. This raises the 
question whether a consistent quantum theory can be formulated. Since the free theory is 
actually on the boundary between stability and instability, the interactions will
decide on which side of the ``stability divide'' the interacting theory will be found \cite{cw3}.
This question depends crucially on the properties of the ground state - unfortunately
a difficult non-perturbative question. As one of the alternatives it was proposed
that a non-perturbative mass for the chiral tensor fields - the chirons - is generated.
This can indeed stabilize the ground state such that a consistent quantum field theory 
is possible. An alternative for the non-perturbative ground state would be a 
non-vanishing vacuum expectation value for the chiral tensor fields. This would imply a 
spontaneous breaking of Lorentz symmetry and rotation symmetry. While this second
alternative may lead to a consistent quantum field theory with a stable ground state,
it is certainly not consistent with observations. In this case chiral tensors cannot be 
used for a solution to the gauge hierarchy problem.

In this paper we study the question of spontaneous Lorentz symmetry breaking via an 
expectation value of the chiral antisymmetric tensor fields $\beta_{\mu\nu}(x)
=-\beta_{\nu\mu}(x)$. We are interested in homogeneus $\beta_{\mu\nu}$. The 
symmetries of the model forbid terms quadratic in $\beta$ - this absence of a local mass 
term is a central ingredient for the solution to the gauge hierarchy problem. The
effective potential $U(\beta)$ for homogeneous $\beta$ therefore depends on invariants
which involve at least four powers of $\beta$. In a polynomial expansion the leading
terms are quartic interactions $\sim\beta^4$. Renormalizable couplings multiply these 
interactions. For suitable choices of the short-distance or ``classical'' couplings the 
quartic ``classical potential'' is positive semidefinite. However, the couplings are scale 
dependent and it may turn out that an instability of the origin $\beta=0$
occurs at lower scales, indicating a nonzero expectation value of $\beta$. In this paper
we investigate the running of the quartic couplings in one loop order. We will
concentrate on one particular coupling which has the property that a negative value would
necessarily lead to instability in the limit of a quartic potential. 
In one-loop order we find that this coupling indeed flows 
towards negative values as the renormalization scale is lowered. We discuss the 
implications for spontaneous Lorentz symmetry breaking in the conclusions.  

\section{Chiral antisymmetric tensor fields} 

The principal idea of the model relies on the observation that chiral information 
can be carried by an antisymmetric tensor field $\beta_{\mu\nu}(x)=-\beta_{\nu\mu}(x)$. 
While the renormalizable interactions between the fermions $\psi$ and the gauge bosons 
preserve a large global flavor symmetry, the observed fermion masses imply that this 
flavor symmetry must be broken. This explicit breaking must be mediated by couplings 
connecting the left- and right-handed fermions $\psi_L$ and $\psi_R$ which belong to 
the representations $(2,1)$ and $(1,2)$ of the Lorentz symmetry. Bilinears involving 
$\psi_L$ and $\bar{\psi}_R$ transform as $(2,1)\otimes(2,1)=(3,1)+(1,1)$ and similar 
for $\psi_R\otimes\bar{\psi}_L=(1,3)+(1,1)$. The violation of the flavor symmetry may 
therefore either involve the Higgs scalar $(1,1)$ or the antisymmetric tensor 
$\beta_{\mu\nu}~(3,1)+(1,3)$. In fact, the six independent components of $\beta_{\mu\nu}$ 
can be decomposed into two inequivalent three dimensional irreducible representations 
of the Lorentz group
\begin{equation}\label{1}
\beta^\pm_{\mu\nu}=\frac{1}{2}\beta_{\mu\nu}\pm\frac{i}{4}\epsilon_{\mu\nu}\ 
^{\kappa\lambda}\beta_{\kappa\lambda},
\end{equation}
with $\epsilon_{\mu\nu\kappa\lambda}$ the totally antisymmetric tensor 
$(\epsilon_{0123}=1)$. We observe that the representations $(3,1)$ and $(1,3)$ 
are complex conjugate to each other. 

The complex fields $\beta^\pm_{\mu\nu}$ should belong to doublets of weak isospin and 
carry hypercharge $Y=1$. The explicit breaking of the flavor symmetry can then be encoded 
in the ``chiral couplings'' $\bar{F}_{U,D,L}$, i.e.
\begin{equation}\label{2}
-{\cal L}_{ch}=\bar{u}_R\bar{F}_U\tilde{\beta}_+q_L-\bar{q}_L
\bar{F}^\dagger_U\bar{\tilde{\beta}}_+ u_R 
+\bar{d}_R\bar{F}_D\bar{\beta}_-q_L-\bar{q}_L\bar{F}^\dagger_D\beta_-d_R 
+\bar{e}_R\bar{F}_L\bar{\beta}_-l_L-\bar{l}_L\bar{F}^\dagger_L\beta_-e_R,
\end{equation}
with
\begin{equation}
 \beta_\pm = \frac{1}{2}\beta^\pm_{\mu\nu}\sigma^{\mu\nu}, \qquad
 \sigma_{\pm}^{\mu\nu}=\frac{i}{4} [\gamma^\mu,\gamma^\nu](1\pm\gamma^5) .
\end{equation}
Details of our notation can be found in \cite{cw1}. The form of the interaction is dictated by
$SU(3)\times SU(2)\times U(1)$ gauge symmetry and Lorentz symmetry. 

Beyond conserved baryon number and lepton number  
the Lagrangian (\ref{2}) is further invariant with 
respect to a discrete $Z_2$-symmetry that we denote by $G_A$. It acts 
\begin{equation}\label{A2a}
G_A(d_R)=-d_R,~G_A(e_R)=-e_R,~G_A(\beta^-)=-\beta^- ,
\end{equation}
while all other fields are invariant. We will require that our model preserves the 
$G_A$-symmetry. The $G_A$-symmetry has an important consequence: no local mass term 
is allowed for the fields $\beta^\pm$! Indeed, a Lorentz singlet is contained 
in $\beta^+\beta^+$ or $\beta^+(\beta^-)^*$, the first being forbidden by hypercharge 
and the second by $G_A$-symmetry.

The absence of an allowed mass term is in sharp contrast to the Higgs mechanism and 
constitutes  a crucial part of the solution to the hierarchy problem. On the other hand, 
there exist kinetic terms $\sim(\beta^+)^*\beta^+$ and $(\beta^-)^*\beta^-$. The most 
general one allowed by all symmetries reads (after a suitable rescaling and for 
vanishing gauge fields)
\begin{equation}\label{No2}
-{\cal L}^{ch}_{\beta,kin}=\frac{1}{4}
\{(\partial^\rho\beta^{\mu\nu})^*\partial_\rho\beta_{\mu\nu}-4
(\partial_\mu\beta^{\mu\nu})^*\partial_\rho\beta^\rho\ _\nu\}.
\end{equation}
In terms of the unconstrained three-component fields $(k,l,j=1\dots 3)$
\begin{equation}\label{No4}
 B^+_k = -i \beta^+_{0k} = \frac{1}{2}\epsilon_{klm}\beta^+_{lm}\quad, \qquad
 B^-_k = i \beta^-_{0k} = \frac{1}{2}\epsilon_{klm}\beta^-_{lm},
\end{equation}
the kinetic term reads in momentum space ($\Omega$: four-volume)
\begin{equation}\label{No5}
-{\cal L}^{ch}_{\beta,kin}=\Omega^{-1}
\int\frac{d^4q}{(2\pi)^4} \{B^{+*}_k(q)P_{kl}(q)B^+_l(q)
 +B^{-*}_k(q)P^*_{kl}(q)B^-_l(q)\},
\end{equation}
with
\begin{equation}\label{No6}
  P_{kl}=-(q^2_0+q_jq_j)\delta_{kl}+2q_kq_l-2i\epsilon_{klj}q_0q_j.
\end{equation}
The inverse propagator $P_{kl}$ obeys the relation $(q^2=q^\mu q_\mu$)
\begin{equation}\label{No7}
  P_{kl}P^*_{lj}=(q_kq_k-q^2_0)^2\delta_{mj}=q^4\delta_{mj}\quad,
\end{equation}
and is invertible for $q^2\neq 0$. The only poles of the propagator 
occur for $q^2=0$, indicating massless excitations. 
Since the fields $B^\pm_k$ are unconstrained and have well defined propagators for 
$q^2\neq 0$, they constitute a convenient basis for the loop calculation in the next 
section. 
The chiral couplings (\ref{2}) can be translated to the basis of fields 
$B^\pm_k$ by using
\begin{eqnarray}\label{31AA}
  \beta_+ = & -2B^+_k\sigma^k_+ ,& \qquad \beta_- = -2B^-_k\sigma^k_-,\nonumber\\
  \bar{\beta}_+ = & -2B^{+*}_k\sigma^k_- ,&\qquad \bar{\beta}_- = -2B^{-*}_k\sigma^k_+,
\end{eqnarray}
with $\sigma^k_\pm$ defined in terms of the Pauli matrices $\tau^k$ as
\begin{equation}\label{31AB}
  \sigma^k_+=\left(\begin{array}{cc}
  \tau^k&0\\0&0\end{array}\right)\ ,\ \sigma^k_-=\left(\begin{array}{cc}
  0&0\\0&\tau^k\end{array}\right),
\end{equation}
in a basis where
\begin{equation}\label{gammas}
 \gamma^\mu =-i\,\left(\begin{array}{cc} 0 & \tau^\mu \\
 \bar{\tau}^\mu & 0 \end{array}\right), \quad
 \tau^\mu = (1,\tau^i), \; \bar{\tau}^\mu = (1,-\tau^i).
\end{equation} 

In terms of $B^\pm_k$ the most general quartic interactions 
allowed by the symmetries read
\begin{eqnarray}\label{QI1}
-{\cal L}_{\beta,4}=U(B^+,B^-)&=&
\frac{\tau_+}{4}\big[(B^+_k)^\dagger B^+_l\big]
\big[(B^+_k)^\dagger B^+_l\big]+(+\rightarrow -)\nonumber\\
&&+\tilde{\tau}_1\big[(B^+_k)^\dagger B^-_k\big]
\big[(B^-_l)^\dagger B^+_l\big]\nonumber\\
&&+\tilde{\tau}_2\big[(B^+_k)^\dagger B^+_l\big]
\big[(B^-_l)^\dagger B^-_k\big]\nonumber\\
&&+\frac{\tau_3}{4}\big[(B^+_k)^\dagger B^-_k\big]
\big[(B^+_l)^\dagger B^-_l\big]+c.c.\nonumber\\
&&+\frac{\tau_4}{4}\big[(B^+_k)^\dagger B^-_l\big]
\big[(B^+_k)^\dagger B^-_l\big]+c.c.\nonumber\\
&&+\frac{\tau_5}{4}\big[(B^+_k)^\dagger B^-_l\big]
\big[(B^+_l)^\dagger B^-_k\big]+c.c. .
\end{eqnarray}
Here [~] indicates the contraction of $SU(2)_L$-indices
of the $B$-doublets. (We have used the tilde in
$\tilde{\tau}_1$ and $\tilde{\tau}_2$ to distinguish them from the couplings $\tau_1$
and $\tau_2$ used in \cite{cw1}, where the corresponding terms were written
in a slightly different way. One has $\tilde{\tau}_1=\tau_1-\tau_2$, 
$\tilde{\tau}_2= 2 \tau_2$.) The couplings $\tau_\pm,\tilde{\tau}_1,\tilde{\tau}_2$ are real 
whereas $\tau_3,\tau_4,\tau_5$ are complex. The quartic interactions (\ref{QI1}) 
are the most general ones consistent with the discrete symmetry 
$G_A$. For
$\tau_3=\tau_4=\tau_5=0$ our model exhibits a further global $U(1)_A$ symmetry.

\section{One-loop four-point functions}

We want to compute the running of the quartic couplings $\tau$. We concentrate here on 
$\tau_+$, which is the only coupling relating four $B^+$ fields and no $B^-$ fields.
If $\tau_+$ runs to negative values one may suspect an instability leading to
Lorentz symmetry breaking. Indeed, we may consider $U(B^+)\equiv U(B^+,B^-=0)$.
If $U(B^+)$ shows a minimum for $B^+_0\neq 0$ one can conclude that the ground state
cannot be at $B^+=0$, $B^-=0$. This would indicate spontaneous Lorentz symmetry breaking 
independently from the precise form of the effective potential $U(B^+,B^-)$.                              
The one-loop four point function of the $B^+$ fields receives two types of 
contributions, the first type involving a fermion loop and four chiral vertices
with coupling $F$ (cf eq.~\ref{2}), the second type involving a $B$-field loop and two
vertices with coupling $\tau_+$, $\tilde{\tau}_1$ or $\tilde{\tau}_2.$ 
For the fermion loop we only retain the largest coupling, $f_t$,
between the chirons and the top quarks. It is this coupling that grows large at the
``chiral scale'' $\Lambda_{ch}$. We ignore
the gauge interactions here, since their contribution can be regarded as small,
at least in the momentum range near $\Lambda_{ch}$ where $f_t$ grows large.
We also ignore the couplings $\tau_3$, $\tau_4$ and $\tau_5$ for simplicity. The corresponding
interactions can be forbidden by requiring the mentioned $U(1)_A$ symmetry.    

In order to study the scale dependence of the couplings we introduce a suitable
infrared cutoff $\propto k$ such that only quantum fluctuations with momenta
$q^2>k^2$ are included. This leads to the concept of the effective average action \cite{cw5}.
All couplings depend now on $k$, and the scale evolution is described by appropriate
$\beta$-functions, i.e.
\begin{equation}
 \frac{\partial}{\partial\ln k}\tau_+(k)=\beta_+(f_t,\tau_\alpha).
\end{equation}
Correspondingly, a polynomial expansion of the effective potential reads
\begin{equation}\label{leff}
 U_k(B^+)=
 \frac{\tau_+(k)}{4}\big[(B^+_k)^\dagger B^+_l\big]\big[(B^+_k)^\dagger B^+_l\big] .
\end{equation}
We denote the two weak isospin components of $B^+_i$ as $B^{++}_i$ and
$B^{+0}_i$, according to their property of being electrically charged or neutral
after electroweak symmetry breaking. Without loss of generality we concentrate on the 
dependence of $U_k$ on $B^{+0}$, taking $B^{++}=0$.

\subsection{Fermion loop}

We assume that the couplings of the $B^+$ fields to the
right-handed $u$ and $c$ quarks are
negligible as compared to $f_t$, the coupling to the right-handed top quark. Note that
$B^+$ does not couple to the right-handed down-type quarks or leptons. 
In the following we compute the diagram with four external fields
$B^{+0}$ with zero momentum,
where two right-handed and two left-handed top quarks
propagate in the loop, as shown in fig.~1. The $B$-field 
indices $j$ and $n$ are summed over.

\begin{figure}
\begin{center}
\begin{picture}(300,120)(0,0)

\DashArrowArcn(80,60)(25,360,270){3} \DashArrowArcn(80,60)(25,270,180){3}
\DashArrowArcn(80,60)(25,180,90){3} \DashArrowArcn(80,60)(25,90,0){3}
\Vertex(105,60){2} \Vertex(55,60){2}
\Vertex(80,85){2} \Vertex(80,35){2}
\ArrowLine(135,60)(105,60) \ArrowLine(25,60)(55,60)
\ArrowLine(80,85)(80,115) \ArrowLine(80,35)(80,5)
\Text(120,65)[bl]{$B_k^{+0}$} \Text(40,65)[br]{$B_k^{+0}$}
\Text(85,100)[l]{$B_l^{+0}$} \Text(85,20)[l]{$B_l^{+0}$}

\DashArrowArcn(220,60)(25,360,270){3} \DashArrowArcn(220,60)(25,270,180){3}
\DashArrowArcn(220,60)(25,180,90){3} \DashArrowArcn(220,60)(25,90,0){3}
\Vertex(245,60){2} \Vertex(195,60){2}
\Vertex(220,85){2} \Vertex(220,35){2}
\ArrowLine(245,60)(275,60) \ArrowLine(165,60)(195,60)
\ArrowLine(220,85)(220,115) \ArrowLine(220,5)(220,35)
\Text(260,65)[bl]{$B_l^{+0}$} \Text(180,65)[br]{$B_k^{+0}$}
\Text(225,100)[l]{$B_l^{+0}$} \Text(225,20)[l]{$B_k^{+0}$}

\end{picture}
\caption{Feynman diagrams with four external $B^{+0}$ fields and a top quark loop.
The second diagram exists only for a non-vanishing top quark mass.}
\end{center}
\end{figure}
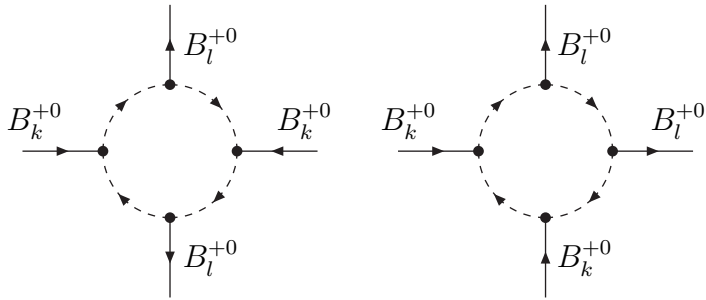

For a computation of $\beta_+$ we can use a rather simple infrared cutoff.
For the fermion loop we employ 
a mass for the top quark $k=m_t$. (This mass is introduced here ``by hand'' and 
violates the $SU(2)$ gauge symmetry. It should not be confounded with the physical
top quark mass which is generated by spontaneous electroweak symmetry breaking
at a scale of the order of $\Lambda_{ch}$. The details of the infrared cutoff do not 
matter here, since the running of the dimensionless coupling $\tau_+$ is logarithmic.)

The expression for the diagram can be either read off directly from the Feynman
rules corresponding to the kinetic and interaction terms, or it can be obtained
by evaluating the top quark loop in an external $B^{+0}$-field. 
We can then extract the fermion loop contribution to $\tau_+$ by taking the 
fourth variation with respect to $B$.
\begin{equation}\label{vary1}
 \tau_+^{(1,f)} = \frac{1}{9 \Omega_4}
 \frac{\partial}{\partial(B_n^{+0})^*}\frac{\partial}{\partial B_j^{+0}}
 \frac{\partial}{\partial(B_n^{+0})^*}\frac{\partial}{\partial B_j^{+0}}
 \;\Gamma_t^{(1l)} \bigg{|}_{B=0},
\end{equation}    
where $\Omega_4$ is the four-volume of integration. The top loop contribution
to the (scale dependent) effective action reads
\begin{equation}\label{g1l}
 \Gamma_t^{(1l)}= i \; Tr \ln P_t.
\end{equation}
Here $P_t$ is the inverse top quark propagator in the presence of background
$B$-fields with zero momentum,
\begin{equation}\label{tprop}
 P_t(q,q')=\left(-\gamma^\mu q_\mu +m_t \gamma^5 + 2 f_t
 \,\sigma_-^k (B_k^{+0})^* -2 f_t \,\sigma_+^k B_k^{+0}
 \right)\delta(q-q').
\end{equation}  
The trace $Tr$ involves a sum over the three
quark colors, an integral over momentum and the trace $tr$ over spinor indices.
The resulting expression is
\begin{eqnarray}
 \tau_+^{(1,f)}  = -\frac{32}{3} i f_t^4 \int \frac{d^4 q}{(2 \pi)^4} tr 
  & \bigg{\{} & \sigma^j_+ \frac{1}{P_t(q)} \sigma^n_- \frac{1}{P_t(q)} 
  \sigma^j_+ \frac{1}{P_t(q)} \sigma^n_- \frac{1}{P_t(q)} \\ \nonumber 
  & + & 2 \sigma^j_+ \frac{1}{P_t(q)} \sigma^j_+ \frac{1}{P_t(q)} 
  \sigma^n_- \frac{1}{P_t(q)} \sigma^n_- \frac{1}{P_t(q)} 
  \bigg{\}}\bigg{|}_{B=0} 
\end{eqnarray}

Inserting the expression for $P_t$, using $\sigma^k_+ \gamma^\mu \sigma^l_+
= \sigma^k_- \gamma^\mu \sigma^l_- =0$ for the second term, 
and $[\sigma_\pm^j, \gamma^5]=0$ and $\sigma_+^j \sigma_-^n =0$ for the first term, we get
\begin{equation}
  \tau_+^{(1,f)} = -\frac{32}{3} i f_t^4 \int \frac{d^4 q}{(2 \pi)^4} 
  \frac{tr \left\{ \sigma^j_+ \gamma^\mu q_\mu \sigma^n_- \gamma^\nu q_\nu
   \sigma^j_+ \gamma^\lambda q_\lambda \sigma^n_- \gamma^\eta q_\eta 
  -2 m_t^2 \sigma_+^j \sigma_+^j \gamma^\mu q_\mu \sigma_-^n \sigma_-^n \gamma^\nu q_\nu
  \right\} }
  {(q^2 +m_t^2)^4}.
\end{equation}
The trace is at best evaluated using the specific representation (\ref{gammas}). 
One gets 
\begin{eqnarray}\nonumber
 tr\{ \cdots \} &=& q_\mu q_\nu q_\lambda q_\eta \; tr \{
 \tau^j \tau^\mu \tau^n \bar{\tau}^\nu \tau^j \tau^\lambda 
 \tau^n \bar{\tau}^\eta \}-18 m_t^2 q^2\\
 &=& q_\mu q_\nu q_\lambda q_\eta \; (2 g^{\mu\nu}g^{\lambda\eta}
  +2 g^{\mu\eta}g^{\nu\lambda} -10 g^{\mu\lambda}g^{\nu\eta}
  -6 i \varepsilon^{\mu\nu\lambda\eta}) -18 m_t^2 q^2 \\ \nonumber
 &=& -6 q^4 -18 m_t^2 q^2 ,
\end{eqnarray}
where we used the following properties of the Pauli matrices:
\begin{equation}
 \{\tau^m ,\tau^n \}=2 \delta_{mn}, \qquad \tau^m \tau^k \tau^m =-\tau^k,
  \qquad tr(\tau^k \tau^m \tau^n) = 2i \varepsilon_{kmn},
\end{equation}
\begin{equation}
 tr(\tau^m \tau^n \tau^p \tau^q) = 2 \delta_{mn}\delta_{pq}
  -2 \delta_{mp}\delta_{nq} +2 \delta_{mq}\delta_{np}.  
\end{equation}
We note the different sign of the trace as compared to a Yukawa theory.
Replacing the coupling to $B^{+0}$ in eq.~(\ref{tprop}) by a corresponding
coupling to the Higgs scalar,
the diagram with four external scalars involves the
trace of $\gamma^\mu \gamma^\nu \gamma^\lambda \gamma^\eta$, i.e. without
the $\sigma_\pm^j$ factors. This trace is $4 g^{\mu\nu}g^{\lambda\eta}
+4 g^{\mu\eta}g^{\nu\lambda} -4 g^{\mu\lambda}g^{\nu\eta}$, leading to
a numerator $+4 q^4$. The resulting contribution to the $\beta$-function
of the quartic scalar coupling $\lambda$ is therefore of opposite sign
as the contribution to $\beta_+$. This sign will play a crucial role.

The integral 
\begin{equation}
 \tau_+^{(1,f)} = \frac{32}{3} i f_t^4 \int \frac{d^4 q}{(2 \pi)^4} 
 \frac{6 q^4 + 18 m_t^2 q^2}{(q^2+m_t^2)^4}
\end{equation}
is logarithmically divergent, but in fact we are interested only in the 
derivative of the integral with respect to $m_t^2$,
\begin{equation}
 \frac{\partial\tau_+^{(1,f)}}{\partial m_t^2}
 = -64 i f_t^4 \int \frac{d^4 q}{(2 \pi)^4} 
 \frac{q^4 + 9 m_t^2 q^2}{(q^2+m_t^2)^5}.
\end{equation}
This integral is easily evaluated using a Wick rotation in momentum space,
$q_0= i q_0^E$. The result is
\begin{equation}
 \frac{\partial\tau_+^{(1,f)}}{\partial m_t^2}
 =\frac{4}{\pi^2}\frac{f_t^4}{m_t^2}.
\end{equation}

Identifying $m_t$ as an infrared cutoff, we conclude that the fermion loop
contributes to the beta function of $\tau_+$ a term
\begin{equation}\label{beta6}
 \frac{\partial \tau_+^{(1,f)}}{\partial\ln k}= 
 \beta_+^{(1,f)}=\frac{8}{\pi^2} f_t^4 .
\end{equation}
The sign of the fermion contribution
to $\beta_+$ is positive, as it was noted in the context of an abelian model
for chiral tensors by Avdeev and Chizhov \cite{ac}.
The positive sign of $\beta_+$ implies that
the fermion loops tend to drive $\tau_+(k)$ to smaller values as $k$ is lowered.
If nothing stops this flow, the coupling $\tau_+$ would turn negative for low
enough $k$. This may be taken as the first indication for a possible instability
associated to spontaneous Lorentz symmetry breaking. The running of the quartic 
chiron interaction therefore contrasts with the quartic scalar interaction,
where the top-loop tends to increase the value of $\lambda$ as $k$ is lowered.
(This effect leads to a lower bound for the mass of the Higgs scalar in the
electroweak standard model.)

\subsection{Chiron loop}

We next need the contribution from loops with internal chiron lines. 
Similar as for eq.~(\ref{vary1}), the 
chiron loops are computed as the fourth variation of
\begin{equation}\label{cl1}
 \Gamma_B^{(1l)}=-\frac{i}{2}Tr \ln S_B^{(2)}
\end{equation}
with respect to the $B^{+0}$ fields. Here $S_B^{(2)}$,
the second variation of the 
bosonic part of the action with respect to the $B$-fields, 
is a $24 \times 24$ matrix,
involving the 12 components of $B_{iap}$ and the 12 components of $B^*_{jbq}$. 
The indices $i,j$ run from 1 to 3, $a,b$ denote chirality ($+$ or $-$) and
$p,q$ denote the weak isospin component ($+$ or $0$). From the kinetic and quartic
terms in the Lagrangian (ignoring $\tau_3$, $\tau_4$ and $\tau_5$) 
we read off $S_B^{(2)}$ (suppressing the momenta in the notation):
\begin{equation}\label{bprop1}
 \frac{\delta^2 S_B}{\delta B^*_{iap}\delta B_{jbq}}=
 P_{ij}\delta_{ab}\delta_{pq} + \tau_+ T^{(+)}_{iap,jbq}+ \tau_- T^{(-)}_{iap,jbq}
 + \tilde{\tau}_1 T^{(1)}_{iap,jbq} + \tilde{\tau}_2 T^{(2)}_{iap,jbq}
\end{equation}
with
\begin{eqnarray}\label{bprop2}
 T^{(+)}_{iap,jbq} &=& \frac{1}{2}\delta_{a+}\delta_{b+}
  (\delta_{pq} B^*_{i+r}B_{j+r} + B^*_{i+q}B_{j+p}) \\ \nonumber
 T^{(-)}_{iap,jbq} &=& \frac{1}{2}\delta_{a-}\delta_{b-}
  (\delta_{pq} B^*_{i-r}B_{j-r} + B^*_{i-q}B_{j-p}) \\ \nonumber
 T^{(1)}_{iap,jbq} &=& \delta_{a+}\delta_{b+}B^*_{j-q}B_{i-p}
  + \delta_{a+}\delta_{b-}\delta_{ij}\delta_{pq} B^*_{k-r}B_{k+r}\\ \nonumber
  &+& \delta_{a-}\delta_{b+}\delta_{ij}\delta_{pq} B^*_{k+r}B_{k-r}
  + \delta_{a-}\delta_{b-}B^*_{j+q}B_{i+p}\\ \nonumber
 T^{(2)}_{iap,jbq} &=& \delta_{a+}\delta_{b+} \delta_{pq} B^*_{j-r}B_{i-r}
  + \delta_{a+}\delta_{b-}\delta_{ij} B^*_{k-q}B_{k+p}\\ \nonumber
  &+& \delta_{a-}\delta_{b+}\delta_{ij} B^*_{k+q}B_{k-p}
  + \delta_{a-}\delta_{b-} \delta_{pq} B^*_{j+r}B_{i+r}.  
\end{eqnarray}
Similarly, one finds
\begin{equation}\label{bprop3}
 \frac{\delta^2 S_B}{\delta B_{iap}\delta B_{jbq}}=
 \tau_+ U^{(+)}_{iap,jbq}+ \tau_- U^{(-)}_{iap,jbq}
 + \tilde{\tau}_1 U^{(1)}_{iap,jbq} + \tilde{\tau}_2 U^{(2)}_{iap,jbq}
\end{equation}
where
\begin{eqnarray}\label{bprop4}
 U^{(+)}_{iap,jbq} &=& \frac{1}{2}\delta_{a+}\delta_{b+}\delta_{ij}
  B^*_{k+p}B^*_{k+q}\\ \nonumber
 U^{(-)}_{iap,jbq} &=& \frac{1}{2}\delta_{a-}\delta_{b-}\delta_{ij}
  B^*_{k-p}B^*_{k-q}\\ \nonumber
 U^{(1)}_{iap,jbq} &=& \delta_{a-}\delta_{b+} B^*_{i+p}B^*_{j-q}
  + \delta_{a+}\delta_{b-} B^*_{i-p}B^*_{j+q}\\ \nonumber
 U^{(2)}_{iap,jbq} &=& \delta_{a-}\delta_{b+} B^*_{i+q}B^*_{j-p}
  + \delta_{a+}\delta_{b-} B^*_{i-q}B^*_{j+p}
\end{eqnarray}
and
\begin{equation}\label{bprop5}
 \frac{\delta^2 S_B}{\delta B^*_{iap}\delta B^*_{jbq}}=
  \left( \frac{\delta^2 S_B}{\delta B_{iap}\delta B_{jbq}} \right)^*.
\end{equation}

In order to make the inverse propagator more explicit, we
specialize to a background with nonvanishing $B_k^{+0}\equiv\bar{B}_k$,
but vanishing $B^-$ and $B^{++}$ fields. In this case the inverse 
propagator does not mix the $B^{++}$, $B^{+0}$, $B^{-+}$ and $B^{-0}$
components. The $24 \times 24$ matrix $S^{(2)}_B$ can therefore be divided into
$6 \times 6$ blocks. For the neutral $B^+$ we use a basis
\begin{equation}
 S^{(2)}_{B+0}=\frac{1}{2} \int_q \left( B_{k+0}^*(q),B_{k+0}(q) \right)
 \tilde{P}_{kl}^{0}(q)\left( B_{l+0}(q),B_{l+0}^*(q) \right) ^T 
\end{equation}
with 
\begin{equation}\label{v1}
 \tilde{P}_{kl}^0 = \left( \begin{array}{ccc}
  P_{kl}+\tau_+ \bar{B}_k^* \bar{B}_l &,& \tau_+ \delta_{kl}\bar{B}_m^* \bar{B}_m \\
  \tau_+ \delta_{kl}\bar{B}_m^* \bar{B}_m &,& P_{kl}^* + \tau_+ \bar{B}_k \bar{B}_l^*
 \end{array} \right) . 
\end{equation}
Similarly we find for the charged $B^+$
\begin{equation}\label{v2}
 \tilde{P}_{kl}^+ = \left( \begin{array}{ccc}
  \tilde{P}_{kl}^{(c)} &,& 0 \\ 0 &,& \tilde{P}_{kl}^{(c)*}
 \end{array}\right), \qquad \
 \tilde{P}_{kl}^{(c)}= P_{kl} + \frac{\tau_+}{2}\bar{B}_k^* \bar{B}_l.
\end{equation}
The $B^-$ propagator can be similarly read off from eqs.~(\ref{bprop1},\ref{bprop2}). As a concrete
example we may choose $\bar{B}_k = \bar{B} \delta_{k3}$, ${\mathcal{B}}\equiv
\tau_+ \bar{B}^2 /2$ and take the momentum in the propagator as 
$\vec{q}=(q_1,0,q_3)$. This yields in the charged sector
\begin{equation}\label{v3}
 \tilde{P}^{(c)}= \left( \begin{array}{ccccc}
  -q_0^2+q_1^2-q_3^2 &,& -2iq_0 q_3 &,& 2q_1q_3 \\
  2iq_0q_3 &,& -q_0^2-q_1^2-q_3^2 &,& -2i q_0 q_1 \\
  2 q_1 q_3 &,& 2i q_0 q_1 &,& -q_0^2-q_1^2+q_3^2 + \mathcal{B}
 \end{array}\right).
\end{equation} 

We may vary eq.~(\ref{cl1})
with respect to the constant neutral $B^{+0}_3$ field (which effectively means
choosing the background as in the example above), and find, after
some calculation, the one-loop chiron contribution
\begin{eqnarray}\nonumber
\tau_+^{(1,b)} & \equiv & \frac{1}{\Omega_4}
 \frac{\partial}{\partial(B_3^{+0})^*}\frac{\partial}{\partial B_3^{+0}}
 \frac{\partial}{\partial(B_3^{+0})^*}\frac{\partial}{\partial B_3^{+0}}
 \;\Gamma_B^{(1l)} \bigg{|}_{B=0}\\ \label{cl2}
&=& i \int \frac{d^4 q}{(2 \pi)^4} q^{-4} \left( 21 \tau_+^2
 + 6(\tilde{\tau}_1 + \tilde{\tau}_2 )^2 + 6\tilde{\tau}_2^2 \right).
\end{eqnarray}
Since a local mass term for the $B$ fields is forbidden by the symmetries,
we have to assume a non-local mass term to generate an effective infrared cutoff.
In the integral (\ref{cl2}), 
the mass is implemented as usual via $q^{-4} \rightarrow (q^2+m^2)^{-2}$. 
One can then evaluate the finite integral for $\partial\tau_+^{(1,b)} /\partial m^2$,
\begin{equation}
 \frac{\partial\tau_+^{(1,b)}}{\partial m^2}
 = \frac{1}{48 \pi^2 m^2} \left( 7 \tau_+^2
 + 2(\tilde{\tau}_1 + \tilde{\tau}_2 )^2 + 2\tilde{\tau}_2^2 \right) .
\end{equation}
Associating $k=m$, this leads to a contribution
\begin{equation}
 \frac{\partial \tau_+^{(1,b)}}{\partial\ln k}= \beta_+^{(1,b)}= \frac{1}{ 24 \pi^2}
 \left( 7 \tau_+^2
 + 2(\tilde{\tau}_1 + \tilde{\tau}_2 )^2 + 2\tilde{\tau}_2^2 \right). 
\end{equation}

The result must be the same when we vary (\ref{cl1}) with repect to other 
components or charges of $B^+$. This is required by the symmetries, but appears
as a non-trivial fact in our $B$-field basis. 
Several diagrams contribute, and the structure of the diagrams for four
external $B_1^{+0}$ fields is very different from, say, those with one $B_1^{+0}$,
one $B_1^{++}$, one $B_2^{+0}$ and one $B_2^{++}$, as shown in figures 2 and 3. 
The diagrams always 
add up such that the contribution from each component is equal, as required
by the symmetry.  

\begin{figure}
\begin{center}
\scalebox{0.8}{
\begin{picture}(400,242)(0,0)

\Vertex(50,210){2} \Vertex(50,160){2}
\ArrowLine(50,210)(20,240) \ArrowLine(50,210)(80,240)
\ArrowLine(20,130)(50,160) \ArrowLine(80,130)(50,160)
\ArrowArc(50,185)(25,270,90) \ArrowArcn(50,185)(25,270,90)
\Text(30,225)[r]{$B_1^{+0}$} \Text(70,225)[l]{$B_1^{+0}$}
\Text(30,145)[r]{$B_1^{+0}$} \Text(70,145)[l]{$B_1^{+0}$} 
\Text(50,208)[t]{$\tau_+/2$} \Text(50,165)[b]{$\tau_+/2$}
\Text(20,185)[r]{$P_{ij}^{+0}$} \Text(80,185)[l]{$P_{ij}^{+0}$} 

\Vertex(150,185){2} \Vertex(200,185){2}
\ArrowLine(150,185)(120,215) \ArrowLine(200,185)(230,215)
\ArrowLine(120,155)(150,185) \ArrowLine(230,155)(200,185)
\ArrowArcn(175,185)(25,180,360) \ArrowArcn(175,185)(25,0,180)
\Text(135,210)[b]{$B_1^{+0}$} \Text(215,210)[b]{$B_1^{+0}$}
\Text(135,160)[t]{$B_1^{+0}$} \Text(215,160)[t]{$B_1^{+0}$} 
\Text(153,185)[l]{$\tau_+$} \Text(198,185)[r]{$\tau_+$}
\Text(175,220)[b]{$P_{11}^{+0}$} \Text(175,155)[t]{$P_{11}^{+0}$} 

\Vertex(290,185){2} \Vertex(340,185){2}
\ArrowLine(290,185)(260,215) \ArrowLine(340,185)(370,215)
\ArrowLine(260,155)(290,185) \ArrowLine(370,155)(340,185)
\ArrowArcn(315,185)(25,180,360) \ArrowArcn(315,185)(25,0,180)
\Text(275,210)[b]{$B_1^{+0}$} \Text(355,210)[b]{$B_1^{+0}$}
\Text(275,160)[t]{$B_1^{+0}$} \Text(355,160)[t]{$B_1^{+0}$} 
\Text(293,185)[l]{$\frac{\tau_+}{2}$} \Text(338,185)[r]{$\frac{\tau_+}{2}$}
\Text(315,220)[b]{$P_{11}^{++}$} \Text(315,155)[t]{$P_{11}^{++}$} 

\Vertex(150,65){2} \Vertex(200,65){2}
\ArrowLine(150,65)(120,95) \ArrowLine(200,65)(230,95)
\ArrowLine(120,35)(150,65) \ArrowLine(230,35)(200,65)
\ArrowArcn(175,65)(25,180,360) \ArrowArcn(175,65)(25,0,180)
\Text(135,90)[b]{$B_1^{+0}$} \Text(215,90)[b]{$B_1^{+0}$}
\Text(135,40)[t]{$B_1^{+0}$} \Text(215,40)[t]{$B_1^{+0}$} 
\Text(146,65)[r]{$\tilde{\tau}_1+\tilde{\tau}_2$} \Text(204,65)[l]{$\tilde{\tau}_1+\tilde{\tau}_2$}
\Text(175,100)[b]{$P_{11}^{-0}$} \Text(175,35)[t]{$P_{11}^{-0}$} 

\Vertex(290,65){2} \Vertex(340,65){2}
\ArrowLine(290,65)(260,95) \ArrowLine(340,65)(370,95)
\ArrowLine(260,35)(290,65) \ArrowLine(370,35)(340,65)
\ArrowArcn(315,65)(25,180,360) \ArrowArcn(315,65)(25,0,180)
\Text(275,90)[b]{$B_1^{+0}$} \Text(355,90)[b]{$B_1^{+0}$}
\Text(275,40)[t]{$B_1^{+0}$} \Text(355,40)[t]{$B_1^{+0}$} 
\Text(295,65)[l]{$\tilde{\tau}_2$} \Text(338,65)[r]{$\tilde{\tau}_2$}
\Text(315,100)[b]{$P_{11}^{-+}$} \Text(315,35)[t]{$P_{11}^{-+}$} 

\end{picture} }
\end{center}
\caption{Feynman diagrams with four ``external'' $B_1^{+0}$ chirons
 and one chiron loop. The first diagram exists for all combinations of $(ij)$.}
\end{figure}
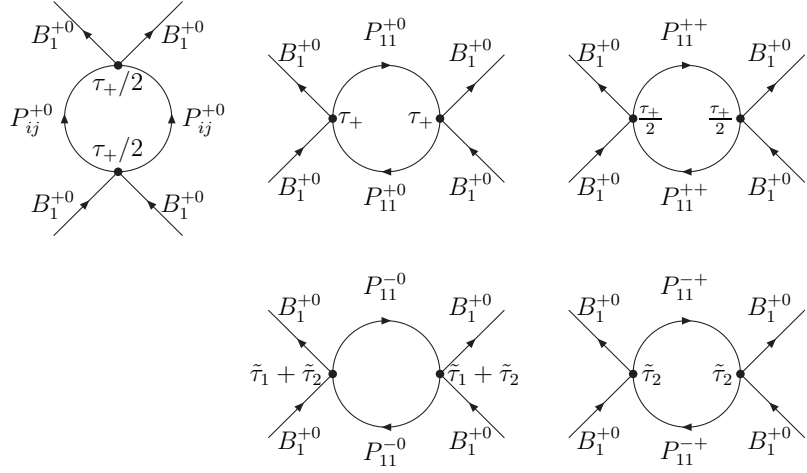

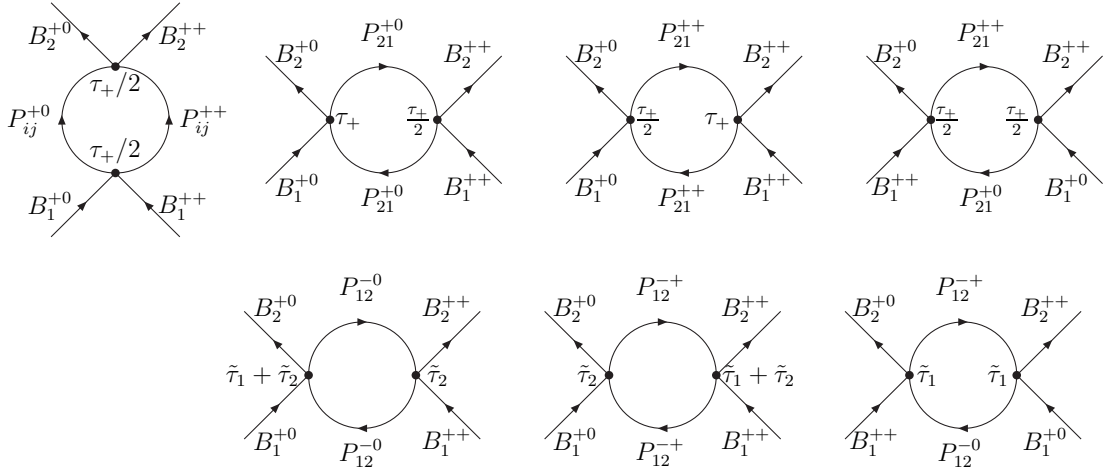
\begin{figure}
\begin{center}
\scalebox{0.8}{
\begin{picture}(600,242)(0,0)

\Vertex(40,210){2} \Vertex(40,160){2}
\ArrowLine(40,210)(10,240) \ArrowLine(40,210)(70,240)
\ArrowLine(10,130)(40,160) \ArrowLine(70,130)(40,160)
\ArrowArc(40,185)(25,270,90) \ArrowArcn(40,185)(25,270,90)
\Text(20,225)[r]{$B_2^{+0}$} \Text(60,225)[l]{$B_2^{++}$}
\Text(20,145)[r]{$B_1^{+0}$} \Text(60,145)[l]{$B_1^{++}$} 
\Text(40,208)[t]{$\tau_+/2$} \Text(40,165)[b]{$\tau_+/2$}
\Text(10,185)[r]{$P_{ij}^{+0}$} \Text(70,185)[l]{$P_{ij}^{++}$} 

\Vertex(140,185){2} \Vertex(190,185){2}
\ArrowLine(140,185)(110,215) \ArrowLine(190,185)(220,215)
\ArrowLine(110,155)(140,185) \ArrowLine(220,155)(190,185)
\ArrowArcn(165,185)(25,180,360) \ArrowArcn(165,185)(25,0,180)
\Text(125,210)[b]{$B_2^{+0}$} \Text(205,210)[b]{$B_2^{++}$}
\Text(125,160)[t]{$B_1^{+0}$} \Text(205,160)[t]{$B_1^{++}$} 
\Text(143,185)[l]{$\tau_+$} \Text(188,185)[r]{$\frac{\tau_+}{2}$}
\Text(165,220)[b]{$P_{21}^{+0}$} \Text(165,155)[t]{$P_{21}^{+0}$} 

\Vertex(280,185){2} \Vertex(330,185){2}
\ArrowLine(280,185)(250,215) \ArrowLine(330,185)(360,215)
\ArrowLine(250,155)(280,185) \ArrowLine(360,155)(330,185)
\ArrowArcn(305,185)(25,180,360) \ArrowArcn(305,185)(25,0,180)
\Text(265,210)[b]{$B_2^{+0}$} \Text(345,210)[b]{$B_2^{++}$}
\Text(265,160)[t]{$B_1^{+0}$} \Text(345,160)[t]{$B_1^{++}$} 
\Text(283,185)[l]{$\frac{\tau_+}{2}$} \Text(328,185)[r]{$\tau_+$}
\Text(305,220)[b]{$P_{21}^{++}$} \Text(305,155)[t]{$P_{21}^{++}$} 

\Vertex(420,185){2} \Vertex(470,185){2}
\ArrowLine(420,185)(390,215) \ArrowLine(470,185)(500,215)
\ArrowLine(390,155)(420,185) \ArrowLine(500,155)(470,185)
\ArrowArcn(445,185)(25,180,360) \ArrowArcn(445,185)(25,0,180)
\Text(405,210)[b]{$B_2^{+0}$} \Text(485,210)[b]{$B_2^{++}$}
\Text(405,160)[t]{$B_1^{++}$} \Text(485,160)[t]{$B_1^{+0}$} 
\Text(423,185)[l]{$\frac{\tau_+}{2}$} \Text(468,185)[r]{$\frac{\tau_+}{2}$}
\Text(445,220)[b]{$P_{21}^{++}$} \Text(445,155)[t]{$P_{21}^{+0}$} 

\Vertex(130,65){2} \Vertex(180,65){2}
\ArrowLine(130,65)(100,95) \ArrowLine(180,65)(210,95)
\ArrowLine(100,35)(130,65) \ArrowLine(210,35)(180,65)
\ArrowArcn(155,65)(25,180,360) \ArrowArcn(155,65)(25,0,180)
\Text(115,90)[b]{$B_2^{+0}$} \Text(195,90)[b]{$B_2^{++}$}
\Text(115,40)[t]{$B_1^{+0}$} \Text(195,40)[t]{$B_1^{++}$} 
\Text(126,65)[r]{$\tilde{\tau}_1+\tilde{\tau}_2$} \Text(186,65)[l]{$\tilde{\tau}_2$}
\Text(155,100)[b]{$P_{12}^{-0}$} \Text(155,35)[t]{$P_{12}^{-0}$} 

\Vertex(270,65){2} \Vertex(320,65){2}
\ArrowLine(270,65)(240,95) \ArrowLine(320,65)(350,95)
\ArrowLine(240,35)(270,65) \ArrowLine(350,35)(320,65)
\ArrowArcn(295,65)(25,180,360) \ArrowArcn(295,65)(25,0,180)
\Text(255,90)[b]{$B_2^{+0}$} \Text(335,90)[b]{$B_2^{++}$}
\Text(255,40)[t]{$B_1^{+0}$} \Text(335,40)[t]{$B_1^{++}$} 
\Text(266,65)[r]{$\tilde{\tau}_2$} \Text(324,65)[l]{$\tilde{\tau}_1+\tilde{\tau}_2$}
\Text(295,100)[b]{$P_{12}^{-+}$} \Text(295,35)[t]{$P_{12}^{-+}$} 

\Vertex(410,65){2} \Vertex(460,65){2}
\ArrowLine(410,65)(380,95) \ArrowLine(460,65)(490,95)
\ArrowLine(380,35)(410,65) \ArrowLine(490,35)(460,65)
\ArrowArcn(435,65)(25,180,360) \ArrowArcn(435,65)(25,0,180)
\Text(395,90)[b]{$B_2^{+0}$} \Text(475,90)[b]{$B_2^{++}$}
\Text(395,40)[t]{$B_1^{++}$} \Text(475,40)[t]{$B_1^{+0}$} 
\Text(415,65)[l]{$\tilde{\tau}_1$} \Text(458,65)[r]{$\tilde{\tau}_1$}
\Text(435,100)[b]{$P_{12}^{-+}$} \Text(435,35)[t]{$P_{12}^{-0}$} 

\end{picture}}
\end{center}
\caption{Feynman diagrams with ``incoming'' $B_1^{+0}$,
$B_1^{++}$, ``outgoing'' $B_2^{+0}$, $B_2^{++}$ chirons
and one chiron loop. The first diagram exists for all combinations of $(ij)$.}
\end{figure}

\section{Running quartic chiron coupling}

So far we have used ``unrenormalized'' or ``bare'' fields for the chirons. 
Since the chiron propagator also receives loop corrections we want to employ 
renormalized fields. Denoting the bare fields of the previous section with a hat,
the renormalized fields $B$ (without hat) are defined as $B=Z_+^{1/2}\hat{B}$,
with $Z_+$ the chiron wave function renormalization. Similarly, we denote the 
bare couplings used in the previous section with a hat, such that our result up
to here can be summarized as
\begin{equation}\label{beta1}
 \frac{\partial}{\partial \ln k}\hat{\tau}_+
 =\frac{1}{\pi^2} \left\{ 8 \hat{f}_t^4 +\frac{7}{24}\hat{\tau}_+^2 
  + \frac{1}{12} \left( (\hat{\tilde{\tau}}_1 + \hat{\tilde{\tau}}_2 )^2 + 
 \hat{\tilde{\tau}}_2^2 \right) \right\}.
\end{equation}
The renormalized couplings obey
\begin{equation}\label{rc}
 f_t^2=\frac{\hat{f}_t^2}{Z_+}, \qquad \tau_++\frac{\hat{\tau}_+}{Z_+^2},
 \qquad \tilde{\tau}_i=\frac{\hat{\tilde{\tau}}_i}{Z_+^2}.
\end{equation}
For the running of the scale dependent renormalized coupling $\tau_+(k)$
this implies 
\begin{equation}\label{beta2}
 \frac{\partial}{\partial \ln k}\tau_+ = 2 \eta_+ \tau_+
 +\frac{1}{Z_+^2}\frac{\partial}{\partial \ln k}\hat{\tau}_+ ,
\end{equation}
where the anomalous dimension has been calculated previously \cite{cw2,cw1},
\begin{equation}\label{eta}
 \eta_+ \equiv - \frac{\partial \ln Z_+}{\partial \ln k}=\frac{f_t^2}{2 \pi^2}.
\end{equation}
Inserting this into eq.~(\ref{beta2}), we get our final one-loop $\beta$-function
\begin{equation}\label{beta3}
 \frac{\partial}{\partial \ln k}\tau_+ = \beta_+
 = \frac{1}{\pi^2}\left\{ f_t^2 \tau_+ + 8 f_t^4 +\frac{7}{24}\tau_+^2 
  + \frac{1}{12} \left( (\tilde{\tau}_1 + \tilde{\tau}_2 )^2 + \tilde{\tau}_2^2
  \right) \right\}.
\end{equation}

All contributions to $\beta_+$ are positive. In consequence, nothing stops the
running of $\tau_+$ towards negative values. 
For a given microscopic value of $\tau_+$, we can obtain an upper bound
for $\tau_+(k)$ if we neglect $\tilde{\tau}_1$, $\tilde{\tau}_2$. In the following
we therefore take $\tilde{\tau}_1 = \tilde{\tau}_2 =0$. A numerical solution 
for $\tau_+(k)$ is shown in fig.~4 for different initial values of $\tau_+$
at the microscopic scale $\Lambda_{UV}$. We associate the latter with some characteristic 
scale for grand unification, $\Lambda_{UV}\sim M_{GUT} \sim 10^{16}GeV$.

\begin{figure}
\begin{center}
\includegraphics{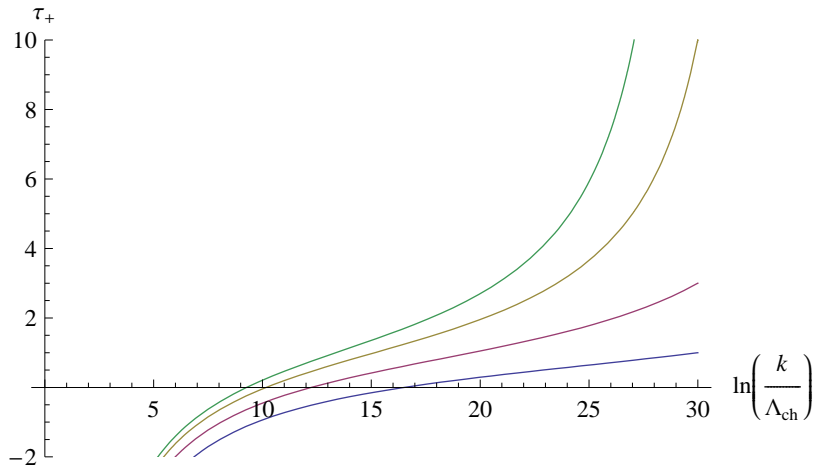}
\caption{Running of the quartic coupling
$\tau_+(k)$ with initial values $\tau_+=1$, 3, 10, 100 at $\Lambda_{UV} \approx 10^{13}
 \Lambda_{ch}\approx e^{30}\Lambda_{ch}$.} 
\end{center}
\end{figure}

For an analytical estimate of the scale where $\tau_+$ turns from positive to negative 
values we only retain the fermion contribution to $\beta_+$. This is justified
for low enough $k$ since  
the chiral freedom property of the coupling $f_t$ implies that $f_t$ grows large.
At the one-loop level, $f_t$ diverges at the scale 
$\Lambda_{ch}$, as computed in ref.~\cite{cw1}, 
\begin{equation}\label{ft}
 f_t^2(k)= \frac{4 \pi^2}{7 \ln (k/\Lambda_{ch})}. 
\end{equation}
As one approaches the scale $\Lambda_{ch}$ from above, the term with $f_t^4$ will
dominate the evolution of $\tau_+$,
\begin{equation}\label{beta4}
 \frac{\partial}{\partial \ln k}\tau_+ \approx \frac{8}{\pi^2}f_t^4
 =\frac{128 \pi^2}{49} \left( \ln \frac{k}{\Lambda_{ch}}\right)^{-2}, 
\end{equation}
with solution
\begin{equation}\label{tauk}
 \tau_+(k)=\tilde{\tau}_+-\frac{128 \pi^2}{49}
 \left( \ln \frac{k}{\Lambda_{ch}}\right)^{-1}.
\end{equation}
If $\tilde{\tau}_+$ is of order 1 or smaller, one concludes that $\tau_+$
becomes negative at a scale roughly 8 orders of magnitude above 
$\Lambda_{ch}$. If we denote by $k_0$ the scale where $\tau_+(k_0)=0$,
one finds
\begin{equation}\label{55a}
 k_0=\Lambda_{ch} \exp\left( \frac{128 \pi^2}{49 \,\tilde{\tau}_+}\right).
\end{equation}
The minimal scale where $\tau_+$ turns negative, corresponding to $\tau_+(\Lambda)
\rightarrow\infty$, is (cf.~fig.~4)
\begin{equation}\label{53a}
  k_{0,min}\approx 10^4 \Lambda_{ch}.
\end{equation}

\section{Spontaneous Lorentz Symmetry Breaking?}
The flow of $\tau_+$ towards negative values may indicate a nonvanishing expectation
value of the chiral tensor field $B$ and therefore a spontaneous breaking of Lorentz
symmetry. In this section we demonstrate that this indeed happens if the top-quark
fluctuations dominate. However, the effects from chiron fluctuations change
qualitatively for a background with nonzero $B_0$: There exist now cubic
vertices with strength $\sim B_0$. Before an inclusion of the chiron fluctuations
is made, which we postpone to future work, no definite conclusion about 
spontaneous Lorentz symmetry breaking can be drawn.

The investigation of Lorentz symmetry breaking proceeds by a computation of the
effective potential $U(B)$. In contrast to the preceeding section, we do not restrict the
discussion to a polynomial expansion around $B=0$, but we rather would like to determine 
the minimum of $U(B)$ which may occur for nonzero $B_0$. If only the top-quark 
fluctuations are taken into account, $U(B)$ can be computed explicitly -
the functional integration over fermion fluctuations is Gaussian. For the top-quark
fluctuations a nonzero $B_0$ acts as an effective infrared cutoff such that no
separate cutoff scale $k$ is needed. 

Let us assume that only the neutral component of $B^+$ takes a constant nonzero value.
Without loss of generality, we may assume that $B^{+0}_k$ obtains its expectation value
in the $x_3$-direction
and define $B \equiv B_3^{+0}$.
We consider the top quark contribution to the effective
potential and rewrite eq.~(\ref{g1l}) in the form
\begin{equation}\label{g1lm}
 U_t(B)= 3i \; \int \frac{d^4 q}{(2 \pi)^4} \ln \det P_t,
\end{equation}
where the determinant acts in the space of the spinor indices.
Inserting $P_t$ from
eq.~(\ref{tprop}) and ignoring the top mass $m_t$ this 
time, the determinant is evaluated as
\begin{eqnarray}\nonumber
 \det P_t(q) &=& \det \left(-\gamma^\mu q_\mu + 2 f_t
 \,\sigma_-^3 B^* -2 f_t \,\sigma_+^3 B \right) \\
 &=& 16 f_t^4 |B|^4 + 8 f_t^2 |B|^2 (-q_0^2 + q_3^2 - q_1^2 -q_2^2) + q^4 .
\end{eqnarray} 
This determinant is not Lorentz invariant, due to the ``wrong sign'' of $q_1^2+q_2^2$. 
We evaluate $U_t$ in Euclidean
momentum coordinates,
\begin{equation}\label{g1le}
 U_t=-3 \int \frac{d^4 q_E}{(2 \pi)^4} \ln \left( 
 16 f_t^4 |B|^4 + 8 f_t^2 |B|^2 (q_{0E}^2 + q_3^2 - q_1^2 -q_2^2) + q_E^4 \right).
\end{equation}
We use radial and angular coordinates
\begin{equation}
 \left( \begin{array}{c} q_{0E} \\ q_3 \end{array}\right) = r_1
 \left( \begin{array}{c} \cos\varphi_1 \\ \sin\varphi_1 \end{array}\right), \qquad
 \left( \begin{array}{c} q_1 \\ q_2 \end{array}\right) =r_2
 \left( \begin{array}{c} \cos\varphi_2 \\ \sin\varphi_2 \end{array}\right),
\end{equation}
\begin{equation}\nonumber
 u=r_1^2-r_2^2 , \qquad v=r_1^2+r_2^2,
\end{equation}
and assume a momentum cutoff at $q_E^2=\Lambda^2 $:
\begin{equation}
 U_t =-\frac{3}{32 \pi^2} \int_0^{\Lambda^2} dv \int_{-v}^v du
 \ln \left( 16 f_t^4 |B|^4 + 8 f_t^2 |B|^2 u + v^2 \right).
\end{equation}
Performing the integral yields
\begin{eqnarray}
 U_t= \frac{1}{32 \pi^2 Y} & \bigg{\{} & 3 \Lambda^4 Y + 2 Y^2 \left( \ln Y
 - \frac{1}{3} \right) \\ \nonumber
 &-& (\Lambda^2 + Y)^3 \left[ \ln (\Lambda^2 +Y)-\frac{1}{3} \right]
 + (\Lambda^2 - Y)^3 \left[ \ln (\Lambda^2 -Y)-\frac{1}{3} \right] \bigg{\}},
\end{eqnarray}
where we employ the abbreviation $Y \equiv 4 f_t^2 |B|^2$. In the regime $Y \ll \Lambda^2$,
and neglecting a constant $\sim\Lambda^4$,  this simplifies to
\begin{equation}
 U_t= \frac{1}{32 \pi^2} \left( 2Y^2 \ln\frac{Y}{\Lambda^2} -\frac{11}{3}Y^2
 \right).
\end{equation}

We wish to compute the minimum of the effective potential
\begin{equation}
 U(|B|^2)=U_0 + U_t = \frac{\tau_{+,\Lambda}}{4}|B|^4 + \frac{f_t^4 |B|^4}{2 \pi^2}
 \left( 2 \ln \frac{|B|^2}{\tilde{\Lambda}^2}-\frac{11}{3} \right),
\end{equation} 
where $\tau_{+,\Lambda}$ is the bare coupling, and $\tilde{\Lambda}=\Lambda/(2f_t)$.
The ultraviolet cutoff is removed by defining a renormalized coupling $\tau_+(\mu)$:
\begin{equation}
 \tau_+(\mu) \equiv 2 \; \frac{\partial^2 U}{\partial (|B|^2)^2}\bigg{|}_{B^2=\mu^2},
\end{equation}
which yields
\begin{equation}\label{epot}
 U(|B|^2)= \frac{\tau_+(\mu)}{4}|B|^4 + \frac{f_t^4 |B|^4}{2 \pi^2}
 \left( 2 \ln \frac{|B|^2}{\mu^2}-3 \right).
\end{equation}
The dependence of $\tau_+$ on the renormalization scale obeys the same renormalization
group equation as eq.~(\ref{beta6}) if we identify $\mu =k$. As it should be, $U$ is 
independent of $\mu$.

The minimum of the effective potential (\ref{epot}) is at $|B|=B_0$, given by
\begin{equation}
 \ln \frac{B_0}{\mu}=\frac{1}{2}-\frac{\pi^2}{8}\frac{\tau_+(\mu)}{f_t^4}.
\end{equation}
We may choose $\mu=\mu_0$ such that $\tau_+(\mu_0)=0$. We conclude that 
$B_0=\exp(\frac{1}{2}) \mu_0$ is of the same order as $\mu_0$. We can compute
$\mu_0$ by replacing $k_0 \rightarrow \mu_0$ in eq.~(\ref{55a}), or, more generally,
by solving eq.~(\ref{beta3}) for given microscopic coupling at $\Lambda_{UV}$.
If the top quark fluctuations dominate, we conclude that the effective 
potential for $B\neq 0$ can take values lower than for $B=0$. In this case the minimum
has to occur for nonzero $B_0$ such that Lorentz symmetry would be spontaneously broken.

\section{Conclusions and Discussion}
We have computed the running of the quartic chiron self-interaction $\tau_+$ in
one loop order. The fluctuation effects drive $\tau_+$ to negative values as
the renormalization scale is lowered. A possible consequence is spontaneous Lorentz
symmetry breaking by a nonzero expectation value of a chiral tensor field.
We have argued that this is indeed the case if the fermion fluctuations dominate.

Before concluding in this direction, however, one has to compute the effects of
chiron fluctuations for $B_0 \neq 0$. These effects may be strongly enhanced
due to the cubic couplings which arise from eq.~(\ref{QI1}) if one of the 
$B$-fields is replaced by the vacuum expectation value $B_0$. It is conceavable 
that these fluctuations have a tendency to drive $B_0$ back to zero.
A possible scenario is a fixed point of the dimensionless ration $B_0(k)/k$,
which would be valid as long as $f_t$ remains in the perturbative regime.

The present computation does not allow us to approach $B_0$ by starting from
large $B$. What is needed is a computation similar to section 5, using 
the propagator in the presence of external $B$-fields, instead of
differentiating first with respect to $B$ and then evaluating the corresponding 
vertex  at $B=0$. Such a calculation is sensitive to the ``Goldstone bosons''
of spontaneously broken Lorentz symmetry. Futhermore, in absence of an additional
infrared cutoff there may be additional instablities for $B_0 \neq 0$. As an
example, let us consider the plane wave solutions for the charged chirons with
inverse propagator (\ref{v3}), and take the momentum direction $\vec{q}=(q_1,0,0)$.
There are three modes with dispersion relation
\begin{equation}
 q_0^2=q_1^2+ \frac{\mathcal{B}}{2}\left\{ 1 \pm \sqrt{1+8q_1^2/\mathcal{B}}\right\}
 , \qquad q_0^2=q_1^2.
\end{equation}
One of them is unstable, with $q_0^2 \approx -q_1^2$ for $q_1^2 \ll \mathcal{B}$.
We infer that a constant nonzero $B^{+0}$ is not likely to be the true ground state.
In view of these findings, the physical implications of the negative quartic
chiron coupling are not yet evident. Nevertheless, we conclude 
that the chiron self-interactions may play an important role for the understanding
of the ground state of chiral tensor theories. It remains to be seen whether 
their running provides an argument for discarding such theories or, in the contrary,
leads to interesting dynamics possibly connected with the generation of a non-local 
mass term for the chirons.

\end{document}